\documentclass{aadebug}

\corr{0309027}{159}

\usepackage{amssymb}

\newtheorem{definition}{Definition}
\newtheorem{theorem}{Theorem}
\newtheorem{lemma}{Lemma}
\newtheorem{corollary}{Corollary}
\newenvironment{proof}{\begin{quote}{\it Proof. }}{\end{quote}}
\newtheorem{example_th}{Example}
\newenvironment{example}{\begin{example_th}\begin{sf}}{\end{sf}\end{example_th}}

\def\domain{{\mathbb{D}}}
\def\var{{\rm var}}

\def\restr#1#2{{#1}|_{{#2}}}
\def\Sol{{\rm Sol}}
\def\downclot{{{\rm CL} \! \downarrow}}
\def\upclot{{{\rm CL} \! \uparrow}}
\def\palm{\texttt{PaLM}}
\def\gnuprolog{\textsc{GNU-Prolog}}

\begin{document}

\runningheads{Gerard Ferrand, Willy Lesaint, Alexandre Tessier}{Towards declarative diagnosis of finite domain constraint programs}

\title{Towards declarative diagnosis of constraint programs over finite domains}

\author{\href{http://www.univ-orleans.fr/SCIENCES/LIFO/Members/ferrand/}{G\'{e}rard~Ferrand}, \href{http://www.univ-orleans.fr/SCIENCES/LIFO/Members/lesaint/}{Willy~Lesaint}, \href{http://www.univ-orleans.fr/SCIENCES/LIFO/Members/tessier/}{Alexandre~Tessier}\addressnum{1}\comma\extranum{1}\comma\extranum{2}}

\address{1}{\href{http://www.univ-orleans.fr/SCIENCES/LIFO/}{Laboratoire d'Informatique Fondamentale d'Orl\'{e}ans}, rue L\'{e}onard de Vinci, BP 6759, F-45067 Orl\'{e}ans Cedex 2, France}

\extra{1}{E-mail: \{\href{mailto:Gerard.Ferrand@lifo.univ-orleans.fr}{Gerard.Ferrand},\href{mailto:Willy.Lesaint@lifo.univ-orleans.fr}{Willy.Lesaint},\href{mailto:Alexandre.Tessier@lifo.univ-orleans.fr}{Alexandre.Tessier}\}@lifo.univ-orleans.fr}
\extra{2}{This work is supported by the French RNTL project \href{http://contraintes.inria.fr/OADymPPaC/}{OADymPPaC} (\href{http://contraintes.inria.fr/OADymPPaC/}{http://contraintes.inria.fr/OADymPPaC/}).}

\pdfinfo{
/Title (Towards declarative diagnosis of constraint programs over finite domains)
/Author (G\'{e}rard Ferrand, Willy Lesaint, Alexandre Tessier)
/Subject (Automated and Algorithmic Debugging of Constraint Programs)
/Keywords (declarative diagnosis; algorithmic debugging; CSP; local consistency operator; fix-point; closure; inductive definition)
}

\begin{abstract}
The paper proposes a theoretical approach of the debugging of constraint programs based on a notion of explanation tree.
The proposed approach is an attempt to adapt algorithmic debugging to constraint programming.
In this theoretical framework for domain reduction, explanations are proof trees explaining value removals.
These proof trees are defined by inductive definitions which express the removals of values as consequences of other value removals.
Explanations may be considered as the essence of constraint programming.
They are a declarative view of the computation trace.
The diagnosis consists in locating an error in an explanation rooted by a symptom.
\end{abstract}


\section{Introduction}

Declarative diagnosis \cite{Shapiro-phd-82} (also known as algorithmic debugging) have been successfully used in different programming paradigms (e.g. in logic programming \cite{Shapiro-phd-82}, in functional programming \cite{FriNil-jfp-94}).
{\em  Declarative} means that the user has no need to consider the computational behavior of the programming system, he only needs a declarative knowledge of the expected properties of the program.
This paper is an attempt to adapt declarative diagnosis to constraint programming thanks to a notion of proof tree called explanation.

Constraint programs are not easy to debug because they are not algorithmic programs \cite{Meier-cp-95} and tracing techniques are limited in usefulness.
Moreover it would be incongruous to use only low level debugging tools whereas for these languages the emphasis is on declarative semantics.
Here we are interested in a wide field of constraint programming: {\em finite domains} and {\em propagation}.

The aim of constraint programming is to solve Constraint Satisfaction Problems (CSP) \cite{Tsang-book-93}, that is to provide an instantiation of the variables which is solution of the constraints.
The solver goes towards the solutions combining two different methods.
The first one (labeling) consists in partitioning the domains.
The second one (domain reduction) reduces the domains eliminating some values which cannot be correct according to the constraints.
In general, the labeling alone is very expensive and domain reduction only provides a superset of the solutions.
Solvers use a combination of these two methods until to obtain singleton domains and test them.

The formalism of domain reduction given in the paper is well-suited to define {\em explanations} for the basic events which are ``the withdrawal of a value from a domain''.
Constraint community is interested in explanations (or nogoods).
An explanation is roughly a set of constraints responsible for a value withdrawal: domain reduction by this set of constraints, or any super-set of it, will always remove this value.
The notions of explanations seem to be an interesting answer to constraint retraction problems.
They have been used and proved useful in many applications such as: dynamic constraint satisfaction problems, over-constrained problems, dynamic backtracking, \ldots
For example, it has been used for failure analysis in \cite{JusOui-wlpe-01} and, in the framework of configuration problems, in \cite{AmiFarMar-ai-02,FFJS-ecai-00}.
See \href{http://www.e-constraints.net}{http://www.e-constraints.net} for more details.
\cite{FerLesTes-entcs-02} details our formalism and our definition of explanation.
It has already permitted to prove the correctness of a large family of constraint retraction algorithms \cite{DFJLOT-flairs-03}.
Here an application to debugging is presented.
There already exists another explanation tree notion defined in \cite{FreLikWal-avcps-00} but it explains solutions obtained by inference in a particular case.
In \cite{FreLikWal-avcps-00} the problem is assumed to have only one solution and the resolution of the problem must not require any search.
The inference rules used to build explanations are defined thanks to cliques of disequalities.
The introduction of labeling in our formalism has already been proposed in \cite{Lesaint-wlpe-02}.
But this introduction complicates the formalism and is not really necessary here because it leads to no conceptual difficulties (labeling can be considered as additional constraints).
The explanations defined in the paper provide us with a declarative view of the computation and their tree structure is used to adapt algorithmic debugging to constraint programming.

From an intuitive viewpoint, we call {\em symptom} the appearance of an anomaly during the execution of a program.
An anomaly is relative to some {\em expected properties} of the program, here to an {\em expected semantics}.
For example, a symptom can be a wrong answer or a missing answer.
This paper focuses on the missing answers.
Symptoms are caused by {\em erroneous constraints}.
Strictly speaking, the localization of an erroneous constraint, when a symptom is given, is {\em error diagnosis}.
It amounts to search for a kind of minimal symptom in the explanation tree.
For our declarative diagnosis approach, the input of a system must include at least (1) the actual {\em program}, (2) the {\em symptom} and (3) a knowledge of the {\em expected} semantics.
This knowledge can be given by the programmer during the diagnosis session or it can be specified by other means but, from a conceptual viewpoint, this knowledge is given by an {\em oracle}.

We are inspired by GNU-Prolog \cite{DiaCod-acm-00}, a constraint programming language over finite domains, because its glass-box approach allows a good understanding of the links between the constraints and the rules used to build explanations.
But this work applies to all solvers over finite domains using propagation whatever the {\em local consistency notion used}.

Section~\ref{sec:preliminaries} defines the basic notions of CSP and program.
In section~\ref{sec:expectedsemantics}, symptoms and errors are described in this framework.
Section~\ref{sec:explanations} defines explanations.
An algorithm for error diagnosis of missing answers is proposed in section~\ref{sec:diagnosis}.

\section{Preliminary notations and definitions}\label{sec:preliminaries}

Our framework uses families instead of cartesian products because it leads to lighter notations.
Indeed, the notion of monotonic operators and least or greatest fixpoints are easier in a set theoretical framework where the order is the set inclusion.

\subsection{Notations}

Let us assume fixed:
\begin{itemize}
  \item a finite set of {\em variable} symbols $V$;
  \item a family $(D_x)_{x \in V}$ where each $D_x$ is a finite non empty set, $D_x$ is the {\em domain} of the variable $x$.
\end{itemize}

We are going to consider various {\em families} $f=(f_i)_{i \in I}$.
Such a family can be identified with the {\em function} $i \mapsto f_i$, itself identified with the {\em set} $\{(i,f_i) \mid i \in I \}$.

In order to have simple and uniform definitions of monotonic operators on a power-set, we use a set which is similar to an Herbrand base in logic programming: we define the {\em domain} by $\domain = \bigcup_{x \in V} (\{x\} \times D_x)$, i.e. $\domain$ is the set of all possible pairs of a variable and its value.

A subset $d$ of $\domain$ is called {\em an environment}.
We denote by $\restr{d}{W}$ the {\em restriction} of $d$ to a set of variables $W \subseteq V$, that is, $\restr{d}{W} = \{ (x,e) \in d \mid x \in W \}$.
Note that, with $d,d' \subseteq \domain$, $d=\bigcup_{x \in V} \restr{d}{\{x\}}$, and $(d \subseteq d' \Leftrightarrow \forall x \in V, \restr{d}{\{x\}} \subseteq \restr{d'}{\{x\}})$.

A {\em tuple} (or {\em valuation}) $t$ is a particular environment such that each variable appears only once: $t \subseteq \domain$ and $\forall x \in V, \exists e \in D_x, \restr{t}{\{x\}}=\{(x,e)\}$.
A {\em tuple} $t$ {\em on} a set of variables $W \subseteq V$, is defined by $t \subseteq \restr{\domain}{W}$ and $\forall x \in W, \exists e \in D_x, \restr{t}{\{x\}}=\{(x,e)\}$.

\subsection{Constraint Satisfaction Problem}

A {\em Constraint Satisfaction Problem} (CSP) on $(V, \domain)$ is made of:
\begin{itemize}
\item a finite set of {\em constraint} symbols $C$;
\item a function $\var : C \rightarrow {\cal P}(V)$, which associates with each constraint symbol the set of variables of the constraint;
\item a family $(T_c)_{c \in C}$ such that: for each $c \in C$, $T_c$ is a set of tuples on $\var(c)$, $T_c$ is the set of {\em solutions} of $c$.
\end{itemize}

From now on, we assume fixed a CSP $(C, \var, (T_c)_{c \in C})$ on $(V, \domain)$.

\begin{definition}
A tuple $t$ is a {\em solution} of the CSP if $\forall c \in C, \restr{t}{\var(c)} \in T_c$.
We denote by $\Sol$ its set of solutions.
\end{definition}

\begin{example} {\em The conference problem \cite{JusOui-wlpe-01}}\\
  Michael, Peter and Alan are organizing a two-day seminar for writing a report on their work.
  In order to be efficient, Peter and Alan need to present their work to Michael and Michael needs to present his work to Alan and Peter.
  So there are four variables, one for each presentation: Michael to Peter (MP), Peter to Michael (PM), Michael to Alan (MA) and Alan to Michael (AM).
  Those presentations are scheduled for a whole half-day each.
  Thus the domains of the variables are $\{1, 2, 3, 4\}$, each value standing for a half-day.

  Michael wants to know what Peter and Alan have done before presenting his own work (MA~$>$~AM, MA~$>$~PM, MP~$>$~AM, MP~$>$~PM).
  Moreover, Michael would prefer not to come the afternoon of the second half-day because he has got a very long ride home (MA $\neq$ 4, MP $\neq$ 4, AM $\neq$ 4, PM $\neq$ 4).
  Finally, note that Peter and Alan cannot present their work to Michael at the same time (AM $\neq$ PM).
  The solutions of this problem are: 

\noindent \{(AM,2),(MA,3),(MP,3),(PM,1)\} and \{(AM,1),(MA,3),(MP,3),(PM,2)\}.

The set of constraints can be written in {\gnuprolog} \cite{DiaCod-acm-00} as:
\begin{verbatim}
conf(AM,MP,PM,MA):-
    fd_domain([MP,PM,MA,AM],1,4),
    MA #> AM, MA #> PM, MP #> AM, MP #> PM,
    MA #\= 4, MP #\= 4, AM #\= 4, PM #\= 4,
    AM #\= PM.
\end{verbatim}
\end{example}

\subsection{Constraint Satisfaction Program}

A program is used to solve a CSP, (i.e to find the solutions) thanks to domain reduction and labeling.
Labeling can be considered as additional constraints, so we concentrate on the domain reduction.
The main idea is quite simple: to remove from the current environment some values which cannot participate to any solution of some constraints, thus of the CSP.
These removals are closely related to a notion of local consistency.
This can be formalized by local consistency operators.

\begin{definition}
A {\em local consistency operator} $r$ is a monotonic function $r: {\cal P}(\domain) \rightarrow {\cal P}(\domain)$.
\end{definition}


\begin{example}
The {\gnuprolog} solver uses local consistency operators following the {\tt X\,in\,r} scheme \cite{CodDia-jlp-96}: for example, {\tt AM in 0..max(MA)-1}.
It means that the values of AM must be between 0 and the maximal value of the environment of MA minus 1.
\end{example}

As we want a contracting operator (i.e. $r(d) \subseteq d$) to reduce the environment, next we will also consider $d \mapsto d \cap r(d)$.
But in general, the local consistency operators are not contracting functions, as shown later to define their dual operators.

A {\em program} on $(V, \domain)$ is a set $R$ of local consistency operators.

\begin{example}
Following the {\tt X~in~r} scheme \cite{CodDia-jlp-96}, the {\gnuprolog} conference problem is implemented by the following program:
\begin{verbatim}
AM in 1..4, MA in 1..4, PM in 1..4, MP in 1..4,
MA in min(AM)+1..infinity, AM in 0..max(MA)-1,
MA in min(PM)+1..infinity, PM in 0..max(MA)-1,
MP in min(AM)+1..infinity, AM in 0..max(MP)-1,
MP in min(PM)+1..infinity, PM in 0..max(MP)-1,
MA in -{4}, AM in -{4}, PM in -{4}, MP in -{4},
AM in -{val(PM)}, PM in -{val(AM)}.
\end{verbatim}
The operator $r$ which corresponds to {\tt X in -\{val(Y)\}} is defined by:
if $\restr{d}{\{{\tt Y}\}}$ is a singleton set then $r(d) = \domain \setminus \restr{d}{\{{\tt Y}\}}$ else $r(d) = \domain$.
\end{example}

From now on, we assume fixed a program $R$ on $(V, \domain)$.

We are interested in particular environments: the common fix-points of the reduction operators $d \mapsto d \cap r(d)$, $r \in R$.
Such an environment $d'$ verifies $\forall r \in R$, $d'=d' \cap r(d')$, that is values cannot be removed according to the operators.

\begin{definition}
Let $r \in R$.
We say an environment $d$ is {\em $r$-consistent} if $d \subseteq r(d)$.

We say an environment $d$ is {\em $R$-consistent} if $\forall r \in R$, $d$ is $r$-consistent.
\end{definition}

Domain reduction from a domain $d$ by $R$ amounts to compute the greatest fix-point of $d$ by $R$.

\begin{definition}\label{def:downwardclosure}
The {\em downward closure} of $d$ by $R$, denoted by $\downclot(d,R)$, is the greatest $d' \subseteq \domain$ such that $d' \subseteq d$ and $d'$ is $R$-consistent.
\end{definition}

In general, we are interested in the closure of $\domain$ by $R$ (the computation starts from $\domain$), but sometimes we would like to express closures of subset of $\domain$ (environments, tuples), for example to take into account dynamic aspects or labeling.

\begin{example}
The closure of the {\gnuprolog} program is:
\begin{center}
\{(AM,1),(AM,2),(MA,2),(MA,3),(MP,2),(MP,3),(PM,1),(PM,2)\}.
\end{center}
It can be computed by a chaotic iteration \cite{FagFowSol-iclp-95,FerLesTes-entcs-02}, but the details about the computation are not in the scope of the paper.
\end{example}

By definition~\ref{def:downwardclosure}, since $d \subseteq \domain$:

\begin{lemma}\label{lem:consistantcloture}
If $d$ is $R$-consistent then $d \subseteq \downclot(\domain, R)$.
\end{lemma}

\subsection{Links between CSP and program}

Of course, the program is linked to the CSP.
The operators are chosen to ``implement'' the CSP.
In practice, this correspondence is expressed by the fact that the program is able to test any valuation.
That is, if all the variables are bounded, the program should be able to answer to the question: ``is this valuation a solution of the CSP ?''.

\begin{definition}\label{def:preserve}
A local consistency operator $r$ {\em preserves the solutions} of a set of constraints $C'$ if, for each tuple $t$, $(\forall c \in C', \restr{t}{\var(c)} \in T_c) \Rightarrow t$ is $r$-consistent.
\end{definition}

In particular, if $C'$ is the set of constraints $C$ of the CSP then we say $r$ preserves the solutions of the CSP.

In the well-known case of arc-consistency, a set of local consistency operators $R_c$ is chosen to implement each constraint $c$ of the CSP.
Of course, each $r \in R_c$ preserves the solutions of $\{c\}$.
It is easy to prove that if $r$ preserves the solutions of $C'$ and $C' \subseteq C$, then $r$ preserves the solutions of $C$.
Therefore $\forall r \in R_c$, $r$ preserves the solutions of the CSP.

Note that an operator may be associated with several constraints, e.g. with path-consistency.

To preserve solutions is a correction property of operators.
A notion of completeness is used to choose the set of operators ``implementing'' a CSP.
It ensures to reject valuations which are not solutions of constraints.
But this notion is not necessary for our purpose.
Indeed, we are only interested in the debugging of missing answers, that is in locating a wrong local consistency operator (i.e. constraints removing too much values).

In the following lemmas, we consider $S \subseteq \Sol$, that is $S$ a set of solutions of the CSP and $\bigcup S$ ($= \bigcup_{t \in S} t$) its projection on $\domain$.

\begin{lemma}\label{lem:preserveconsistant}
Let $S \subseteq \Sol$, if $r$ preserves the solutions of the CSP then $\bigcup S$ is $r$-consistent.
\end{lemma}

\begin{proof}
$\forall t \in S, t \subseteq r(t)$ so $\bigcup S \subseteq \bigcup_{t \in S} r(t)$.
Now, $\forall t \in S, t \subseteq \bigcup S$ so $\forall t \in S, r(t) \subseteq r(\bigcup S)$.
\end{proof}

Extending definition~\ref{def:preserve}, we say $R$ preserves the solutions of $C$ if for each $r \in R$, $r$ preserves the solutions of $C$.

From now on, we consider that the fixed program $R$ preserves the solutions of the fixed CSP.

\begin{lemma}
If $S \subseteq \Sol$ then $\bigcup S \subseteq \downclot(\domain,R)$.
\end{lemma}

\begin{proof}
by lemmas~\ref{lem:consistantcloture} and~\ref{lem:preserveconsistant}.
\end{proof}

Finally, the following corollary emphasizes the link between the CSP and the program.

\begin{corollary}
$\bigcup \Sol \subseteq \downclot(\domain,R)$.
\end{corollary}

The downward closure is a superset (an ``approximation'') of $\bigcup \Sol$ which is itself the projection (an ``approximation'') of $\Sol$.
But the downward closure is the most accurate set which can be computed using a set of local consistency operators in the framework of domain reduction without splitting the domain (without search tree).

\section{Expected Semantics}\label{sec:expectedsemantics}

To debug a constraint program, the programmer must have a knowledge of the problem.
If he does not have such a knowledge, he cannot say something is wrong in his program!
Because constraint programming activity is declarative, this knowledge is declarative.

\subsection{Correctness of a CSP}

At first, the expected semantics of the CSP is considered as a set of tuples: the {\em expected solutions}.
Note that the only relation between the fixed CSP and the fixed program $R$ is that $R$ preserves the solutions of the CSP.
Next definition is motivated by the debugging of missing answer.

\begin{definition}
Let $S$ be a set of tuples.
The CSP is {\em correct} wrt $S$ if $S \subseteq \Sol$.
\end{definition}

Note that if the user exactly knows $S$ then it could be sufficient to test each tuple of $S$ on each local consistency operator or constraint.
But in practice, the user only needs to know some members of $\bigcup S$ and some members of $\domain \setminus \bigcup S$.
We consider the {\em expected environment} $\bigcup S$, that is the approximation of $S$.

By lemma~\ref{lem:preserveconsistant}:

\begin{lemma}
If the CSP is correct wrt a set of tuples $S$ then $\bigcup S$ is $R$-consistent.
\end{lemma}


\subsection{Symptom and Error}

From the notion of expected environment, we can define a notion of symptom.
A symptom emphasizes a difference between what is expected and what is actually computed.

\begin{definition}
$h \in \domain$ is a {\em symptom} wrt an expected environment $d$ if $h \in d \setminus \downclot(\domain,R)$.
\end{definition}

It is important to note that here a symptom is a symptom of missing solution (an expected member of $\domain$ is not in the closure).

\begin{example}\label{exa:symptom}
From now on, let us consider the new following CSP in {\gnuprolog}:
\begin{verbatim}
conf(AM,MP,PM,MA):-
    fd_domain([MP,PM,MA,AM],1,4),
    MA #> AM, MA #> PM, MP #> AM, PM #> MP,
    MA #\= 4, MP #\= 4, AM #\= 4, PM #\= 4,
    AM #\= PM.
\end{verbatim}
As we know, a solution of the conference problem contains (AM,1).
But, the execution provides an empty closure.
So, in particular, (AM,1) has been removed.
Thus, (AM,1) is a symptom.
\end{example}

\begin{definition}
$R$ is {\em approximately correct} wrt $d$ if $d \subseteq \downclot(\domain,R)$.
\end{definition}

Note that $R$ is approximately correct wrt $d$ is equivalent to there is no symptom wrt $d$.
By this definition and lemma~\ref{lem:consistantcloture} we have:

\begin{lemma}
If $d$ is $R$-consistent then $R$ is approximately correct wrt $d$.
\end{lemma}

In other words, if $d$ is $R$-consistent then there is no symptom wrt $d$.
But, our purpose is debugging (and not program validation), so:

\begin{corollary}
Let $S$ be a set of expected tuples.
If $R$ is not approximately correct wrt $\bigcup S$ then $\bigcup S$ is not $R$-consistent, thus the CSP is not correct wrt $S$.
\end{corollary}

The lack of an expected value is caused by an error in the program, more precisely a local consistency operator.
If an environment $d$ is not $R$-consistent, then there exists an operator $r \in R$ such that $d$ is not $r$-consistent.

\begin{definition}
A local consistency operator $r \in R$ is an {\em erroneous operator} wrt $d$ if $d \not\subseteq r(d)$.
\end{definition}

Note that $d$ is $R$-consistent is equivalent to there is no erroneous operator wrt $d$ in $R$.

\begin{theorem}\label{the:symptomerror}
If there exists a symptom wrt $d$ then there exists an erroneous operator wrt $d$ (the converse does not hold).
\end{theorem}

When the program is $R = \bigcup_{c \in C} R_c$ with each $R_c$ a set of local consistency operators preserving the solutions of $c$, if $r \in R_c$ is an erroneous operator wrt $\bigcup S$ then it is possible to say that $c$ is an erroneous constraint.
Indeed, there exists a value $(x,e) \in \bigcup S \setminus r(\bigcup S)$, that is there exists $t \in S$ such that $(x,e) \in t \setminus r(t)$.
So $t$ is not $r$-consistent, so $\restr{t}{\var(c)} \not \in T_c$ i.e. $c$ rejects an expected solution.

\section{Explanations}\label{sec:explanations}

The previous theorem shows that when there exists a symptom there exists an erroneous operator.
The goal of error diagnosis is to locate such an operator from a symptom.
To this aim we now define explanations of value removals. An explanation is a proof tree of a value removal (\cite{FerLesTes-entcs-02} gives more details about explanations).
If a value has been wrongly removed then there is something wrong in the proof of its removal, that is in its explanation.

\subsection{Explanations}

First we need some notations.
Let $\overline{d} = \domain \setminus d$.
In order to help the understanding, we always use the notation $\overline{d}$ for a subset of $\domain$ if intuitively it denotes a set of removed values.

\begin{definition}\label{Def:dualoperator}
Let $r$ be an operator, we denote by $\widetilde{r}$ the {\em dual} of $r$ defined by: $\forall d \subseteq \domain, \widetilde{r}(\overline{d}) = \overline{r(d)}$.
\end{definition}

Definition~\ref{Def:dualoperator} provides a dual view of domain reduction: instead of speaking about values that are kept in the environments this dual view consider the values removed from the environments.

We consider the set of dual operators of $R$:
let $\widetilde{R} = \{ \widetilde{r} \mid r \in R \}$.

\begin{definition}
The {\em upward closure} of $\overline{d}$ by $\widetilde{R}$, denoted by $\upclot(\overline{d}, \widetilde{R})$ exists and is the least $\overline{d'}$ such that $\overline{d} \subseteq \overline{d'}$ and $\forall r \in R$, $\widetilde{r}(\overline{d'}) \subseteq \overline{d'}$.
\end{definition}

Next lemma establishes the correspondence between downward closure of local consistency operators and upward closure of their duals.

\begin{lemma}
$\upclot(\overline{d}, \widetilde{R}) = \overline{\downclot(d, R)}$.
\end{lemma}

\begin{proof}
\begin{tabular}[t]{rcl}
  $\upclot(\overline{d}, \widetilde{R})$ & $=$ & $\mbox{min} \{\overline{d'} \mid \overline{d} \subseteq \overline{d'}, \forall \widetilde{r} \in \widetilde{R}, \widetilde{r}(\overline{d'}) \subseteq \overline{d'} \}$\\
  & $=$ & $\mbox{min} \{\overline{d'} \mid \overline{d} \subseteq \overline{d'}, \forall r \in R, d' \subseteq r(d') \}$\\
  & $=$ & $\overline{\mbox{max}} \{ d' \mid d' \subseteq d, \forall r \in R, d' \subseteq r(d') \}$ \\
\end{tabular}
\end{proof}

In particular, $\upclot(\emptyset, \widetilde{R}) = \overline{\downclot(\domain, R)}$ is the set of values removed by the program during the computation.

Now, we associate rules in the sense of \cite{Aczel-handbook-77} with these dual operators.
These rules are natural to build the complementary of an environment and well suited to provide proof (trees) of value removals.

\begin{definition}
A {\em deduction rule} is a rule $h \leftarrow B$ such that $h \in \domain$ and $B \subseteq \domain$.
\end{definition}

Intuitively, a deduction rule $h \leftarrow B$ can be understood as follows: if all the elements of $B$ are removed from the environment, then $h$ can be removed.

A very simple case is arc-consistency where $B$ corresponds to the well-known notion of support of $h$.
But in general (even for hyper arc-consistency) the rules are more intricate.
Note that these rules are only a theoretical tool to define explanations and to justify the error diagnosis method.
But in practice, this set does not need to be given.
The rules are hidden in the algorithms which implement the solver.

For each operator $r \in R$, we denote by ${\cal R}_r$ a set of deduction rules which defines $\widetilde{r}$, that is, ${\cal R}_r$ is such that: $\widetilde{r} (\overline{d}) = \{ h \in \domain \mid \exists B \subseteq \overline{d}, h \leftarrow B \in {\cal R}_r \}$.
For each operator, this set of deduction rules exists.
There possibly exists many such sets, but for classical notions of local consistency one is always natural \cite{FerLesTes-entcs-02}.
The deduction rules clearly appear inside the algorithms of the solver.
In \cite{AptMon-cp-99} the proposed solver is directly something similar to the set of rules (it is not exactly a set of deduction rules because the heads of the rules do not have the same shape that the elements of the body).

\begin{example}
With the {\gnuprolog} operator {\tt AM in 0..max(MA)-1} are associated the deduction rules (considering that the domains of the variables are $\{1,2,3,4\}$):
\begin{itemize}
\item (AM,1) $\leftarrow$ (MA,2), (MA,3), (MA,4)
\item (AM,2) $\leftarrow$ (MA,3), (MA,4)
\item (AM,3) $\leftarrow$ (MA,4)
\item (AM,4) $\leftarrow \emptyset$
\end{itemize}
Indeed, for the first one, the value 1 is removed from the environment of AM only when the values 2, 3 and 4 are not in the environment of MA.
\end{example}

With the deduction rules, we have a notion of proof tree \cite{Aczel-handbook-77}.
We consider the set of all the deduction rules for all the local consistency operators of $R$: let ${\cal R} = \bigcup_{r \in R} {\cal R}_r$.

We denote by ${{\rm cons}}(h,T)$ the tree defined by: $h$ is the label of its root and $T$ the set of its sub-trees.
The label of the root of a tree $t$ is denoted by ${{\rm root}}(t)$.

\begin{definition}
An {\em explanation} is a proof tree ${{\rm cons}}(h,T)$ with respect to ${\cal R}$; it is inductively defined by: $T$ is a set of explanations with respect to ${\cal R}$ and  $(h \leftarrow \{{{\rm root}}(t) \mid t \in T \}) \in {\cal R}$.
\end{definition}

\begin{example}

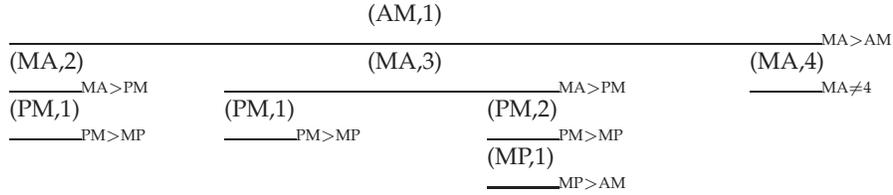
\begin{figure}[t]
\setlength{\unitlength}{3947sp}%
\begingroup\makeatletter\ifx\SetFigFont\undefined%
\gdef\SetFigFont#1#2#3#4#5{%
  \reset@font\fontsize{#1}{#2pt}%
  \fontfamily{#3}\fontseries{#4}\fontshape{#5}%
  \selectfont}%
\fi\endgroup%
\centerline{
\begin{picture}(5124,1171)(139,-673)
\thinlines
{\put(151,239){\line( 1, 0){5100}}}
{\put(151,-61){\line( 1, 0){450}}}
{\put(1501,-61){\line( 1, 0){2100}}}
{\put(4801,-61){\line( 1, 0){450}}}
{\put(3151,-361){\line( 1, 0){450}}}
{\put(151,-361){\line( 1, 0){450}}}
{\put(1501,-361){\line( 1, 0){450}}}
{\put(3151,-661){\line( 1, 0){450}}}
\put(2401,389){\makebox(0,0)[lb]{\smash{\SetFigFont{9}{10.8}{\rmdefault}{\mddefault}{\updefault}{(AM,1)}}}}
\put(2401, 89){\makebox(0,0)[lb]{\smash{\SetFigFont{9}{10.8}{\rmdefault}{\mddefault}{\updefault}{(MA,3)}}}}
\put(4801, 89){\makebox(0,0)[lb]{\smash{\SetFigFont{9}{10.8}{\rmdefault}{\mddefault}{\updefault}{(MA,4)}}}}
\put(151, 89){\makebox(0,0)[lb]{\smash{\SetFigFont{9}{10.8}{\rmdefault}{\mddefault}{\updefault}{(MA,2)}}}}
\put(151,-211){\makebox(0,0)[lb]{\smash{\SetFigFont{9}{10.8}{\rmdefault}{\mddefault}{\updefault}{(PM,1)}}}}
\put(1501,-211){\makebox(0,0)[lb]{\smash{\SetFigFont{9}{10.8}{\rmdefault}{\mddefault}{\updefault}{(PM,1)}}}}
\put(3151,-211){\makebox(0,0)[lb]{\smash{\SetFigFont{9}{10.8}{\rmdefault}{\mddefault}{\updefault}{(PM,2)}}}}
\put(3151,-511){\makebox(0,0)[lb]{\smash{\SetFigFont{9}{10.8}{\rmdefault}{\mddefault}{\updefault}{(MP,1)}}}}
\put(3601,-61){\makebox(0,0)[lb]{\smash{\SetFigFont{6}{8.4}{\rmdefault}{\mddefault}{\updefault}{MA$>$PM}}}}
\put(601,-61){\makebox(0,0)[lb]{\smash{\SetFigFont{6}{8.4}{\rmdefault}{\mddefault}{\updefault}{MA$>$PM}}}}
\put(5251,-61){\makebox(0,0)[lb]{\smash{\SetFigFont{6}{8.4}{\rmdefault}{\mddefault}{\updefault}{MA$\neq$4}}}}
\put(3601,-361){\makebox(0,0)[lb]{\smash{\SetFigFont{6}{8.4}{\rmdefault}{\mddefault}{\updefault}{PM$>$MP}}}}
\put(601,-361){\makebox(0,0)[lb]{\smash{\SetFigFont{6}{8.4}{\rmdefault}{\mddefault}{\updefault}{PM$>$MP}}}}
\put(1951,-361){\makebox(0,0)[lb]{\smash{\SetFigFont{6}{8.4}{\rmdefault}{\mddefault}{\updefault}{PM$>$MP}}}}
\put(3601,-661){\makebox(0,0)[lb]{\smash{\SetFigFont{6}{8.4}{\rmdefault}{\mddefault}{\updefault}{MP$>$AM}}}}
\put(5251,239){\makebox(0,0)[lb]{\smash{\SetFigFont{6}{8.4}{\rmdefault}{\mddefault}{\updefault}{MA$>$AM}}}}
\end{picture}}
\caption{An explanation for (AM,1)}
\label{Fig:expl}
\end{figure}

The explanation of figure~\ref{Fig:expl} is an explanation for (AM,1).
Note that the root (AM,1) of the explanation is linked to its children by the deduction rule (AM,1) $\leftarrow$ (MA,2), (MA,3), (MA,4).
Here to help understanding, since each rule is associated with an operator which is itself associated with a constraint (arc-consistency case), the  constraint is written at the right of the rule.
\end{example}

Finally we prove that the elements removed from the domain are the roots of the explanations.

\begin{theorem}
$\overline{\downclot(\domain,R)}$ is the set of the roots of explanations with respect to ${\cal R}$.
\end{theorem}

\begin{proof}
Let $E$ the set of the roots of explanations wrt to ${\cal R}$.
By induction on explanations $E \subseteq \mbox{min} \{\overline{d} \mid \forall \widetilde{r} \in \widetilde{R}, \widetilde{r}(\overline{d}) \subseteq \overline{d} \}$.
It is easy to check that $\widetilde{r}(E) \subseteq E$.
Hence, $\mbox{min} \{\overline{d} \mid \forall \widetilde{r} \in \widetilde{R}, \widetilde{r}(\overline{d}) \subseteq \overline{d} \} \subseteq E$.
So $E=\upclot(\emptyset,\widetilde{R})$.
\end{proof}

In \cite{FerLesTes-entcs-02} there is a more general result which establishes the link between the closure of an environment $d$ and the roots of explanations of ${\cal R} \cup \{ h \leftarrow \emptyset \mid h \in \overline{d} \}$.
But here, to be lighter, the previous theorem is sufficient because we do not consider dynamic aspects.
All the results are easily adaptable when the starting environment is $d \subset \domain$.

\subsection{Computed explanations}

Note that for error diagnosis, we only need a program, an expected semantics, a symptom and an explanation for this symptom.
Iterations are briefly mentioned here only to understand how explanations are computed in concrete terms, as in the {\palm} system \cite{BarJus-trics-00}.
For more details see \cite{FerLesTes-entcs-02}.

$\downclot(\domain,R)$ can be computed by {\em chaotic iterations} introduced for this aim in \cite{FagFowSol-iclp-95}.

The principle of a chaotic iteration \cite{Apt-tcs-99} is to apply the operators one after the other in a ``fairly'' way, that is such that no operator is forgotten.
In practice this can be implemented thanks to a propagation queue.
Since $\subseteq$ is a well-founded ordering (i.e.~$\domain$ is a finite set), every chaotic iteration is stationary.
The well-known result of confluence \cite{CouCou-saipl-77,FagFowSol-iclp-95} ensures that the limit of every chaotic iteration of the set of local consistency operators $R$ is the {\em downward closure} of $\domain$ by $R$.
So in practice the computation ends when a common fix-point is reached.
Moreover, implementations of solvers use various strategies in order to determine the order of invocation of the operators.
These strategies are used to optimize the computation, but this is out of the scope of this paper.

We are interested in the explanations which are ``computed'' by chaotic iterations, that is the explanations which can be deduced from the computation of the closure.
A chaotic iteration amounts to apply operators one after the other, that is to apply sets of deduction rules one after another.
So, the idea of the incremental algorithm \cite{FerLesTes-entcs-02} is the following:
each time an element $h$ is removed from the environment by a deduction rule $h \leftarrow B$, an explanation is built.
Its root is $h$ and its sub-trees are the explanations rooted by the elements of $B$.

Note that the chaotic iteration can be seen as the trace of the computation, whereas the computed explanations are a declarative vision of it.

The important result is that $\overline{\downclot(\domain,R)}$ is the set of roots of computed explanations.
Thus, since a symptom belongs to $\overline{\downclot(\domain,R)}$, there always exists a computed explanation for each symptom.

\section{Error Diagnosis}\label{sec:diagnosis}

If there exists a symptom then there exists an erroneous operator.
Moreover, for each symptom an explanation can be obtained from the computation.
This section describes how to locate an erroneous operator from a symptom and its explanation.

\subsection{From Symptom to Error}

\begin{definition}
A rule $h \leftarrow B \in {\cal R}_r$ is an {\em erroneous rule} wrt $d$ if $B \cap d = \emptyset$ and $h \in d$.
\end{definition}

Intuitively, an erroneous rule $h \leftarrow B$ can be understood as follows: all the elements of $B$ can be removed from the environment, but $h$ should not be removed.

It is easy to prove that $r$ is an erroneous operator wrt $d$ if and only if there exists an erroneous rule $h \leftarrow B \in {\cal R}_r$ wrt $d$.
Consequently, theorem~\ref{the:symptomerror} can be extended into the next lemma.

\begin{lemma}
If there exists a symptom wrt $d$ then there exists an erroneous rule wrt $d$.
\end{lemma}

We say a node of an explanation is a {\em symptom} wrt $d$ if its label is a symptom wrt $d$.
Since, for each symptom $h$, there exists an explanation whose root is labeled by $h$, it is possible to deal with minimality according to the relation parent/child in an explanation.

\begin{definition}
A symptom is {\em minimal} wrt $d$ if none of its children is a symptom wrt $d$.
\end{definition}

Note that if $h$ is a minimal symptom wrt $d$ then $h \in d$ and the set of its children $B$ is such that $B \subseteq \overline{d}$.
In other words $h \leftarrow B$ is an erroneous rule wrt $d$.

\begin{theorem}
In an explanation rooted by a symptom wrt $d$, there exists at least one minimal symptom wrt $d$ and the rule which links the minimal symptom to its children is an erroneous rule.
\end{theorem}

\begin{proof}
Since explanations are finite trees, the relation parent/child is well-founded.
\end{proof}

To sum up, with a minimal symptom is associated an erroneous rule, itself associated with an erroneous operator.
Moreover, an operator is associated with a constraint (e.g. the usual case of hyper arc-consistency), or a set of constraints.
Consequently, the search for some erroneous constraints in the CSP can be done by the search for a minimal symptom in an explanation rooted by a symptom.

\subsection{Diagnosis Algorithms}

Let $(x,e)$ be a symptom.
Let $E$ be the computed explanation of $(x,e)$.
The aim is to find a minimal symptom in $E$.

A quite simple error diagnosis algorithm for the symptom $(x,e)$ is to ask the user with questions as: ``is $(y,f)$ expected ?'' (i.e. is this removal an anomaly?) until to locate a minimal symptom.

Note that different strategies can be used.
For example, the ``divide and conquer'' strategy: if $n$ is the number of nodes of $E$ then the number of questions is {\cal O}$(log(n))$, that is not much according to the size of the explanation and so not very much compared to the size of the iteration.

\begin{example}
  Let us consider the {\gnuprolog} CSP of example~\ref{exa:symptom}.
  Remind us that its closure is empty whereas the user expects (AM,1) to belong to a solution.
  Let the explanation of figure~\ref{Fig:expl} be the computed explanation of (AM,1).
  A diagnosis session can then be done using this explanation to find the erroneous operator or constraint of the CSP.
  
  Following the ``divide and conquer'' strategy, first question is: {\it ``Is (MA,3) a symptom ?''}.
  According to the conference problem, the knowledge on MA is that Michael wants to know other works before presenting is own work (that is MA$>$2) and Michael cannot stay the last half-day (that is MA is not 4).
  Then, the user's answer is: {\it yes}.
  
  Second question is: {\it ``Is (PM,2) a symptom ?''}.
  According to the conference problem, Michael wants to know what Peter has done before presenting his own work to Alan, so the user considers that (PM,2) belongs to the expected environment: its answer is {\it yes}.

  Third question is: {\it ``Is (MP,1) a symptom ?''}.
  This means that Michael presents his work to Peter before Peter presents his work to him.
  This is contradicting the conference problem: the user answers {\it no}.

  So, (PM,2) is a minimal symptom and the rule (PM,2) $\leftarrow$ (MP,1) is an erroneous one.
  This rule is associated with the operator {\tt PM in min(MP)+1..infinite}, associated with the constraint PM$>$MP.
  Indeed, Michael wants to know what Peter have done before presenting his own work would be written PM$<$MP.

  Note that the user has to answer to only three questions whereas the explanation contains height nodes, there are sixteen removed values and eighteen operators for this problem.
  There exists other stategies not detailed here and ``divide and conquer'' is always a good stategy in practice (about log(number of nodes) questions).
  So, declarative diagnosis seems an efficient way to find an error.
\end{example}

Note that it is not necessary for the user to exactly know the set of solutions, nor a precise approximation of them.
The expected semantics is theoretically considered as a partition of $\domain$: the elements which are expected and the elements which are not.
For the error diagnosis, the oracle only have to answer to some questions (he has to reveal step by step a part of the expected semantics).
The expected semantics can then be considered as three sets: a set of elements which are expected, a set of elements which are not expected and some other elements for which the user does not know.
It is only necessary for the user to answer to the questions.

It is also possible to consider that the user does not answer to some questions, but in this case there is no guarantee to find an error \cite{FerTes-discipl-00}.
Without such a tool, the user is in front of a chaotic iteration, that is a wide list of events.
In these conditions, it seems easier to find an error in the code of the program than to find an error in this wide trace.
Even if the user is not able to answer to the questions, he has an explanation for the symptom which contains a subset of the CSP constraints.

\section{Conclusion}

Our theoretical foundations of domain reduction have permitted to define notions of expected semantics, symptom and error.

Explanation trees provide us with a declarative view of the computation and their tree structure is used to adapt algorithmic debugging \cite{Shapiro-phd-82} to constraint programming.
The proposed approach consists in comparing expected semantics (what the user wants to obtain) with the actual semantics (the closure computed by the solver).
Here, a symptom, which expresses a difference between the two semantics is a missing element, that is an expected element which is not in the closure.
Since the symptom is not in the closure there exists an explanation for it (a proof of its removal).
The diagnosis amounts to search for a minimal symptom in the explanation (rooted by the symptom), that is to locate the error from the symptom.
The traversal of the tree is done thanks to an interaction with an oracle (usually the user): it consists in questions to know if an element is member of the expected semantics.

It is important to note that the user does not need to understand the computation of the constraint solver, unlike a method based on a presentation of the trace.
A declarative approach is then more convenient for constraint programs.
Especially as the user only has a declarative knowledge of its problem/program and the solver computation is too intricate to understand.

The contribution of this paper is the full presentation of the theoretical bases of declarative debugging for constraint programming over finite domains.
Now it remains to implement a declarative diagnoser in order to test the approach on real constraint problems.
Indeed, without an implementation, it is not possible to manage problems other than toy problems because the number of events in the trace becomes quickly huge (for example, magic squares 4x4: 49776 events, 8-queens: 3416 events, magic lists of length 7: 132552 events).
So it is not possible to claim that this approach will help for non trivial constraint problems.
But we think that this kind of diagnosis is well suited because the size of computations is often huge and without such a methodology an errorenous operator cannot be located in a trace.


\newcommand{\etalchar}[1]{$^{#1}$}


\end{document}